\documentclass[twoside,slac_one]{revtex4}
\usepackage{graphicx}
\usepackage{fancyhdr}
\usepackage{amsmath} % American Mathematics Society standards
\usepackage{bm}% bold math
\usepackage{amsxtra}
\usepackage{amssymb}
\usepackage{amsthm}
\usepackage{latexsym}
\usepackage{lscape}

\def\beq{\begin{equation}}
\def\eeq{\end{equation}}
\def\beqa{\begin{eqnarray}}
\def\eeqa{\end{eqnarray}}

\begin{document}

\title{Two-loop corrections to $W$ and $Z$ boson production at high $p_T$}


\author{Nikolaos Kidonakis$^a$ and Richard J. Gonsalves$^b$}
\affiliation{${}^a$ Kennesaw State University, Physics \#1202, Kennesaw, GA 30144, USA}
\affiliation{${}^b$ Department of Physics, University at Buffalo, The State University of New York, Buffalo, NY 14260-1500, USA}

\begin{abstract}
We present new results for the complete two-loop corrections in the 
soft approximation for $W$ and $Z$ boson production at large transverse 
momentum. Analytical expressions for the NNLO approximate corrections are 
used to calculate transverse momentum distributions. Results for the $W$ boson 
$p_T$ distribution  at Tevatron and LHC energies are presented.
\end{abstract}

\maketitle

\thispagestyle{fancy}


%%%%%%%%%%%%%%%%%%%%%%%%%%%%%%%%%%
\section{Introduction}

The production of $W$ and $Z$ bosons in hadron colliders is important in 
testing the Standard Model and in the search for new physics, as it is 
a background to Higgs production and new gauge bosons.
Here we present new results for the theoretical calculation of the 
differential cross section at large transverse momentum.
We begin in Section 2 by presenting the leading-order (LO) partonic production 
channels and presenting some numerical LO results for the $W$ $p_T$ 
distribution.
In Section 3 we discuss the structure of the NLO corrections and identify 
an important subset, the soft-gluon corrections.
In Section 4, we discuss NNLL resummation of these corrections via 
two-loop calculations.
In Section 5 we present the approximate NNLO $p_T$ distribution
for $W$ production at the LHC and the Tevatron.
We conclude in Section 6 with a summary.

\section{Partonic channels at LO}
 
The LO partonic processes for $W$ (or $Z$) production at large $p_T$ are 
$$q(p_a) + g(p_b) \longrightarrow W(Q) + q(p_c)$$
and
$$q(p_a) + {\bar q}(p_b) \longrightarrow W(Q) + g(p_c) \, .$$
We define $s=(p_a+p_b)^2$, $t=(p_a-Q)^2$, $u=(p_b-Q)^2$, 
and $s_4=s+t+u-Q^2$.
At the partonic threshold $s_4 \rightarrow 0$.
As we will see later, the  soft-gluon corrections are of the form 
$[\ln^l(s_4/p_T^2)/s_4]_+$ while the 
virtual corrections are $\delta(s_4)$ terms.

\begin{figure}[h]
\centering
\includegraphics[width=100mm]{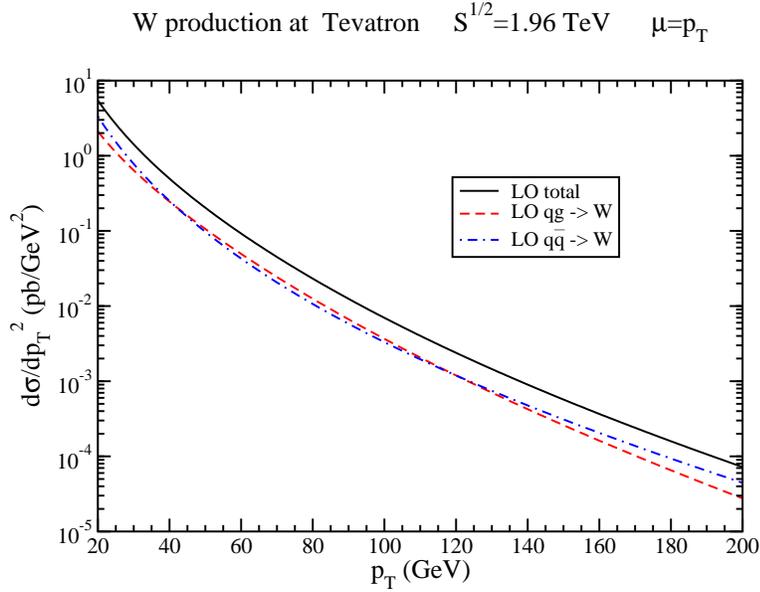}
\caption{LO $p_T$ distribution for $W$ production at the Tevatron.}
\label{Wtevborn}
\end{figure}

In all numerical results below we use the MSTW2008 NNLO pdfs \cite{MSTW}.
We begin with leading-order results for $W$ production at the Tevatron. 
In Fig. \ref{Wtevborn} we show results for the LO 
$p_T$ distribution, $d\sigma/dp_T^2$, 
for $W$ production at the Tevatron. Here we have set the factorization 
scale $\mu_F$ equal to the renormalization scale $\mu_R$, and in the 
numerical results in Fig. \ref{Wtevborn} we set this 
common scale, denoted as $\mu$, equal to $p_T$.
We show separate results for the two LO channels as well as for their sum.
We note that the $qg$ and $q{\bar q}$ channels are  equally important. The
$p_T$ distribution falls rapidly as $p_T$ increases.

\begin{figure}[h]
\centering
\includegraphics[width=100mm]{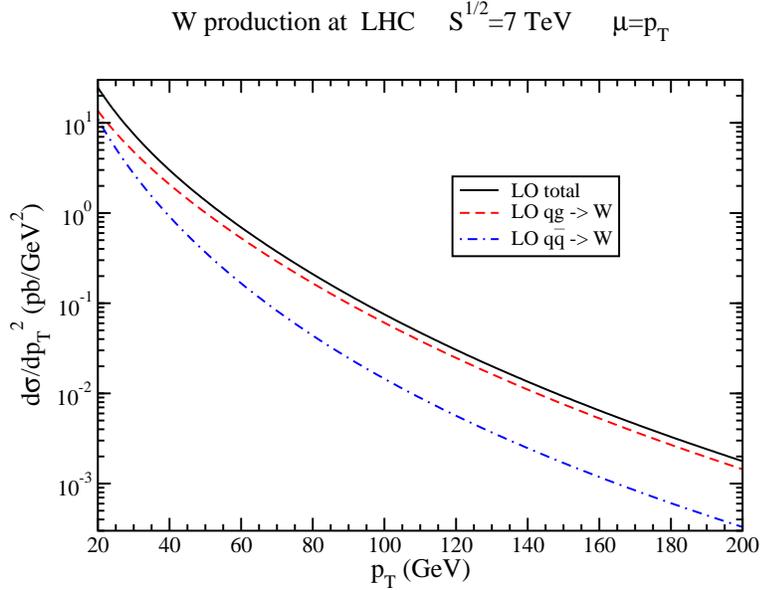}
\caption{LO $p_T$ distribution for $W$ production at the LHC at 7 TeV.}
\label{W7lhcborn}
\end{figure}

Leading-order results for $W$ production at the LHC at 7 TeV energy are 
shown in Fig. \ref{W7lhcborn}, again with $\mu=p_T$.
The $qg$ channel is numerically dominant at this energy for all $p_T$ 
values shown in the plot. 

\begin{figure}[h]
\centering
\includegraphics[width=100mm]{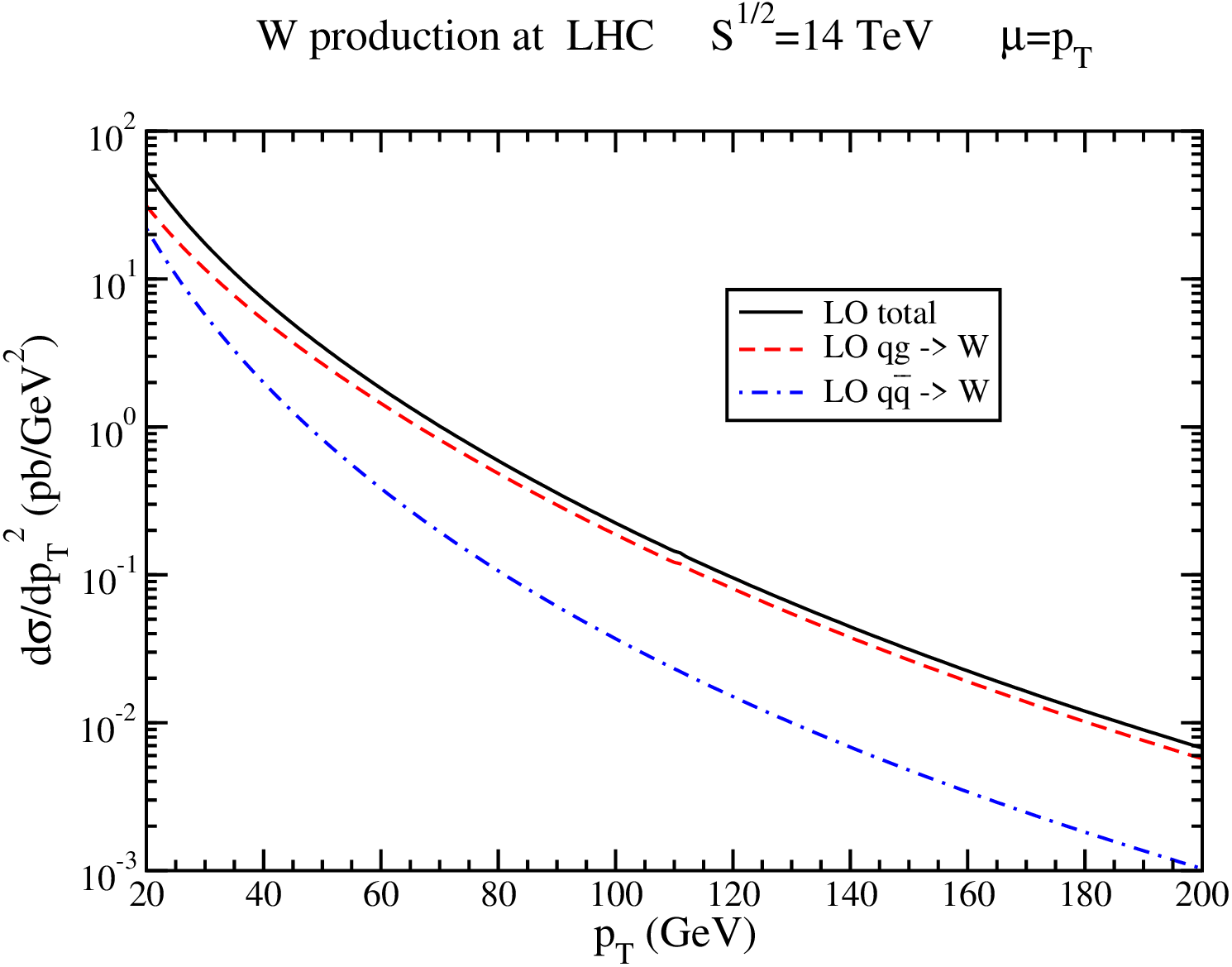}
\caption{LO $p_T$ distribution for $W$ production at the LHC at 14 TeV.}
\label{W14lhcborn}
\end{figure}

The corresponding leading-order results for $W$ production at the LHC 
at 14 TeV energy are shown in Fig. \ref{W14lhcborn}.
Again the $qg$ channel is numerically dominant.

\begin{figure}[h]
\centering
\includegraphics[width=100mm]{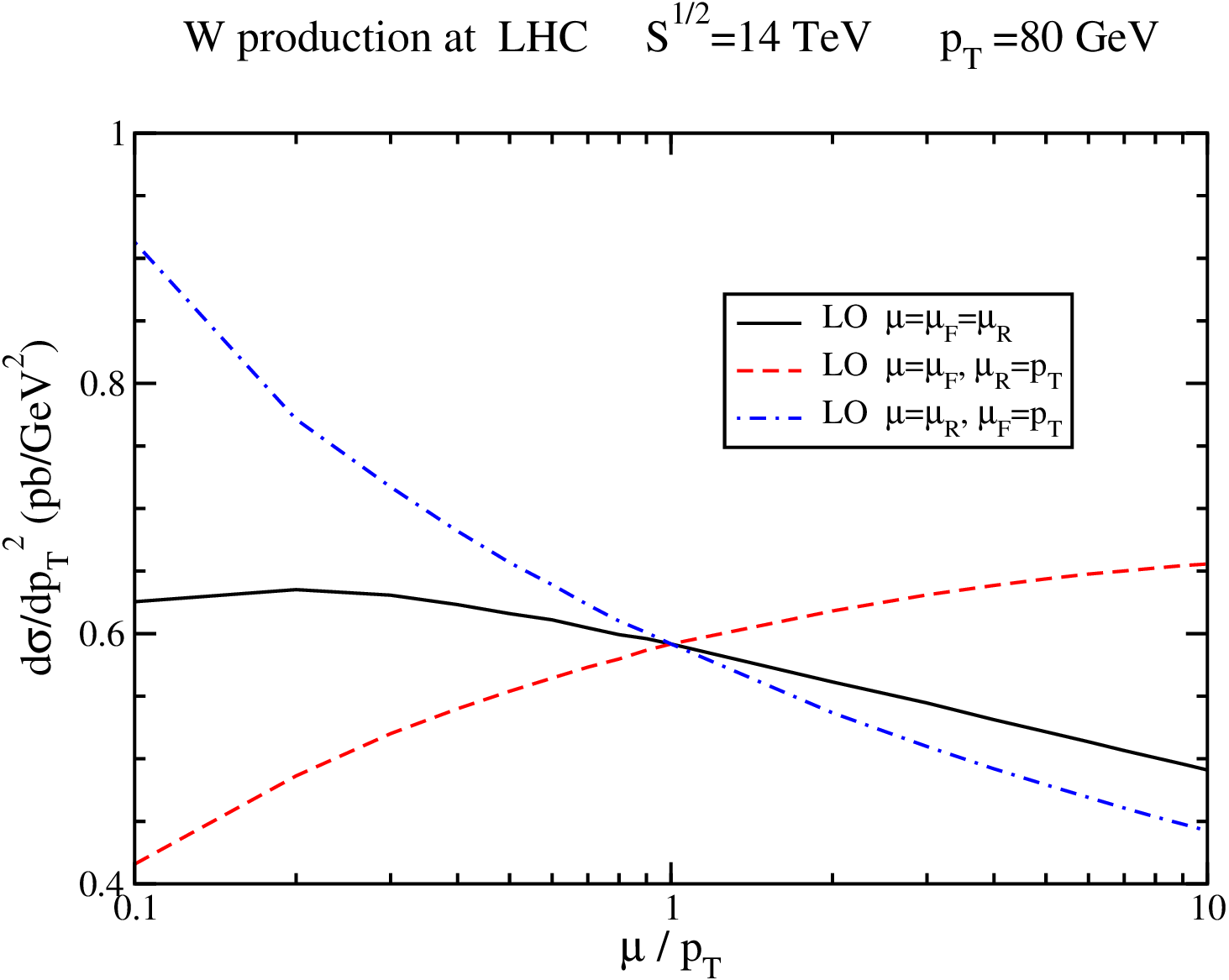}
\caption{Scale dependence of the LO $p_T$ distribution for $W$ production 
at the LHC at 14 TeV.}
\label{W14lhcmuborn}
\end{figure}

In Fig. \ref{W14lhcmuborn} we show the LO scale dependence for $W$ production 
at the LHC at 14 TeV. We have chosen $p_T=80$ GeV and plot the $p_T$ 
distribution as a function of $\mu/p_T$. We show three curves. 
One has the common scale $\mu=\mu_F=\mu_R$ varied by two orders of magnitude. 
Another sets $\mu=\mu_F$ and varies this scale while keeping $\mu_R$ fixed, 
$\mu_R=p_T$. The third sets $\mu=\mu_R$ and varies this scale while fixing 
$\mu_F=p_T$.
At LO the factorization scale, $\mu_F$, and renormalization scale, $\mu_R$, 
dependence largely cancel each other, so that the $\mu=\mu_F=\mu_R$ line shows 
relatively mild variation.

\begin{figure}[h]
\centering
\includegraphics[width=100mm]{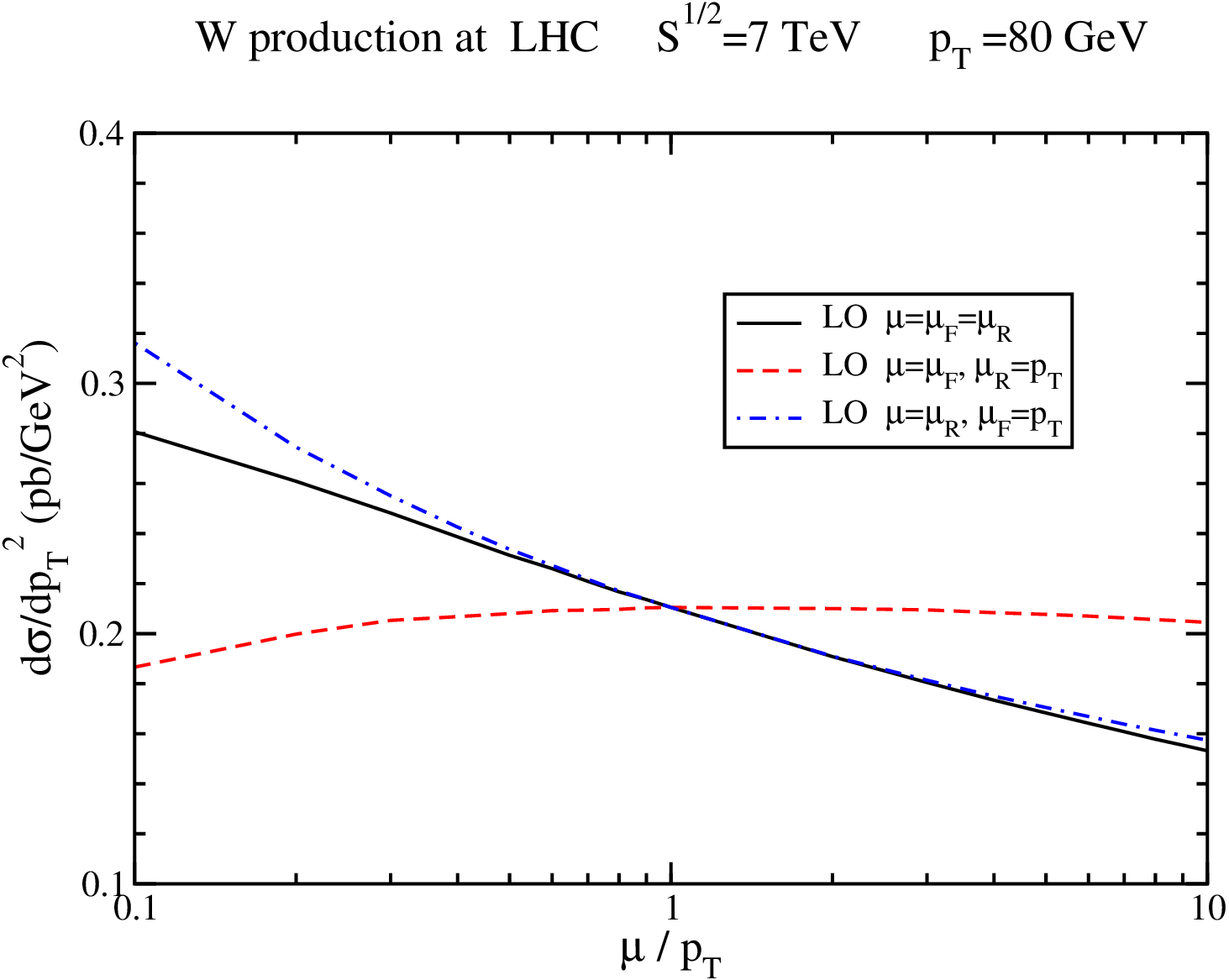}
\caption{Scale dependence of the LO $p_T$ distribution for $W$ production 
at the LHC at 7 TeV.}
\label{W7lhcmuborn}
\end{figure}

In Fig. \ref{W7lhcmuborn} we show the corresponding LO scale dependence 
at 7 TeV energy at the LHC. Comparings Figs. \ref{W14lhcmuborn} and 
\ref{W7lhcmuborn} we note a somewhat different scale  dependence 
at 7 and 14 TeV.

\section{NLO corrections}

The complete NLO corrections were derived in \cite{AR,gpw}.
The NLO cross section can be written as 
\beqa
E_Q\,\frac{d\hat{\sigma}_{f_af_b{\rightarrow}W(Q)+X}}{d^3Q}&=&
\delta(s_4)\alpha_s(\mu_R^2)\left[A(s,t,u) 
+\alpha_s(\mu_R^2) B(s,t,u,\mu_R)\right] + \alpha_s^2(\mu_R^2)C(s,t,u,s_4,\mu_F) \, .
\nonumber 
\eeqa

The coefficient functions $A$, $B$, and $C$ depend on the parton flavors. 
The coefficient $A(s,t,u)$ arises from the LO processes.  
$B(s,t,u,\mu_R)$ is the sum of virtual corrections and of singular terms 
${\sim}\delta(s_4)$ in the real radiative corrections.   
$C(s,t,u,s_4,\mu_F)$ is from real emission processes away from $s_4=0$. 

The NLO corrections are crucial in reducing theoretical uncertainties and 
thus making more meaningful comparisons with experimental data for 
$W$ production \cite{CDF-W,D0-W,ATLAS-W} and $Z$ production 
\cite{CDF-Z,D0-Z,ATLAS-Z} at large transverse momentum.

At NLO and higher orders terms appear that arise from soft-gluon emission.
These soft-gluon corrections are of the form
$$
{\cal D}_l(s_4)\equiv\left[\frac{\ln^l(s_4/p_T^2)}{s_4}\right]_+
$$
and are numerically dominant for processes near partonic threshold, such as 
$W$ or $Z$ production at large $p_T$.
For the order $\alpha_s^n$ corrections $l\le 2n-1$.
At NLO, we have ${\cal D}_1(s_4)$ and ${\cal D}_0(s_4)$ terms, which appear 
in the analytical expressions for the exact NLO corrections.
At NNLO, we have ${\cal D}_3(s_4)$, ${\cal D}_2(s_4)$, ${\cal D}_1(s_4)$, and
${\cal D}_0(s_4)$ terms. 

\section{Two-loop soft-gluon resummation}

We can formally resum these soft logarithms for $W$ and $Z$ production at 
large $p_T$ to all orders in $\alpha_s$.
This resummation was derived at NLL accuracy in \cite{NKVD} and has been applied 
to $W$ production at the Tevatron \cite{NKASV} and the LHC at 14 TeV \cite{GKS}.
In \cite{NKASV} approximate NNLO cross sections were derived 
from the NLL resummed cross section. After matching with the exact NLO corrections, 
the NNLO ${\cal D}_3(s_4)$, ${\cal D}_2(s_4)$, and ${\cal D}_1(s_4)$ terms 
were completely determined.
However, only an approximate form of the NNLO ${\cal D}_0(s_4)$ terms was provided.
New two-loop results for the soft anomalous dimensions for $W$ and 
$Z$ production \cite{NKDISDPF} now allow us to 
determine the NNLO ${\cal D}_0(s_4)$ terms fully. 
We note that recently a different resummation approach, based on Soft 
Collinear Effective Theory, has also been applied to $W$ and $Z$ production 
at large $p_T$ \cite{BLS}.

Soft-gluon resummation follows from factorization properties of the 
cross section, performed in moment space.
The resummed cross section is
\beqa
{\hat{\sigma}}^{res}(N) &=&
\exp\left[ \sum_i E_i(N_i)\right] \, \exp\left[ E'_j(N')\right]\;
\exp \left[\sum_{i=1,2} 2 \int_{\mu_F}^{\sqrt{s}} \frac{d\mu}{\mu}\;
\gamma_{i/i}\left({\tilde N}_i, \alpha_s(\mu)\right)\right] \;
\nonumber\\ && \hspace{-20mm} \times \,
H\left(\alpha_s\right) \;
S \left(\alpha_s\left(\frac{\sqrt{s}}{\tilde N'}\right)\right) \;
\exp \left[\int_{\sqrt{s}}^{{\sqrt{s}}/{\tilde N'}}
\frac{d\mu}{\mu}\; 2\, {\rm Re} \Gamma_S\left(\alpha_s(\mu)\right)\right] 
\nonumber 
\eeqa
where the first exponential resums 
the collinear and soft-gluon radiation from the inital-state partons, 
the second exponential resums corresponding terms from the final state,
$H$ is the hard-scattering function, $S$ is the soft-gluon function and 
$\Gamma_S$ is the soft anomalous dimension.

We expand $\Gamma_S$ as
$$
\Gamma_S=\frac{\alpha_s}{\pi}\Gamma_S^{(1)}
+\frac{\alpha_s^2}{\pi^2}\Gamma_S^{(2)}
+\cdots
$$
The one-loop result, $\Gamma_S^{(1)}$, is obtained from the UV poles 
of one-loop eikonal diagrams, and was first derived in \cite{NKVD}.

\begin{figure}[h]
\centering
\includegraphics[width=80mm]{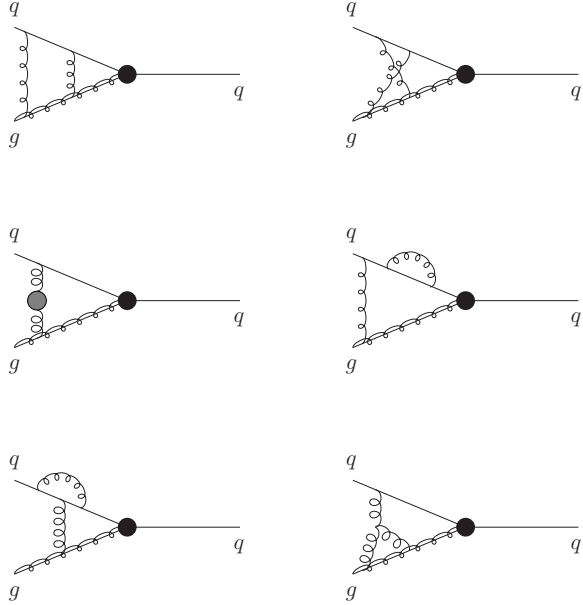}
\caption{Two-loop eikonal diagrams for $W$ or $Z$ production-Set 1.}
\label{W2loop1}
\end{figure}

\begin{figure}[h]
\centering
\includegraphics[width=80mm]{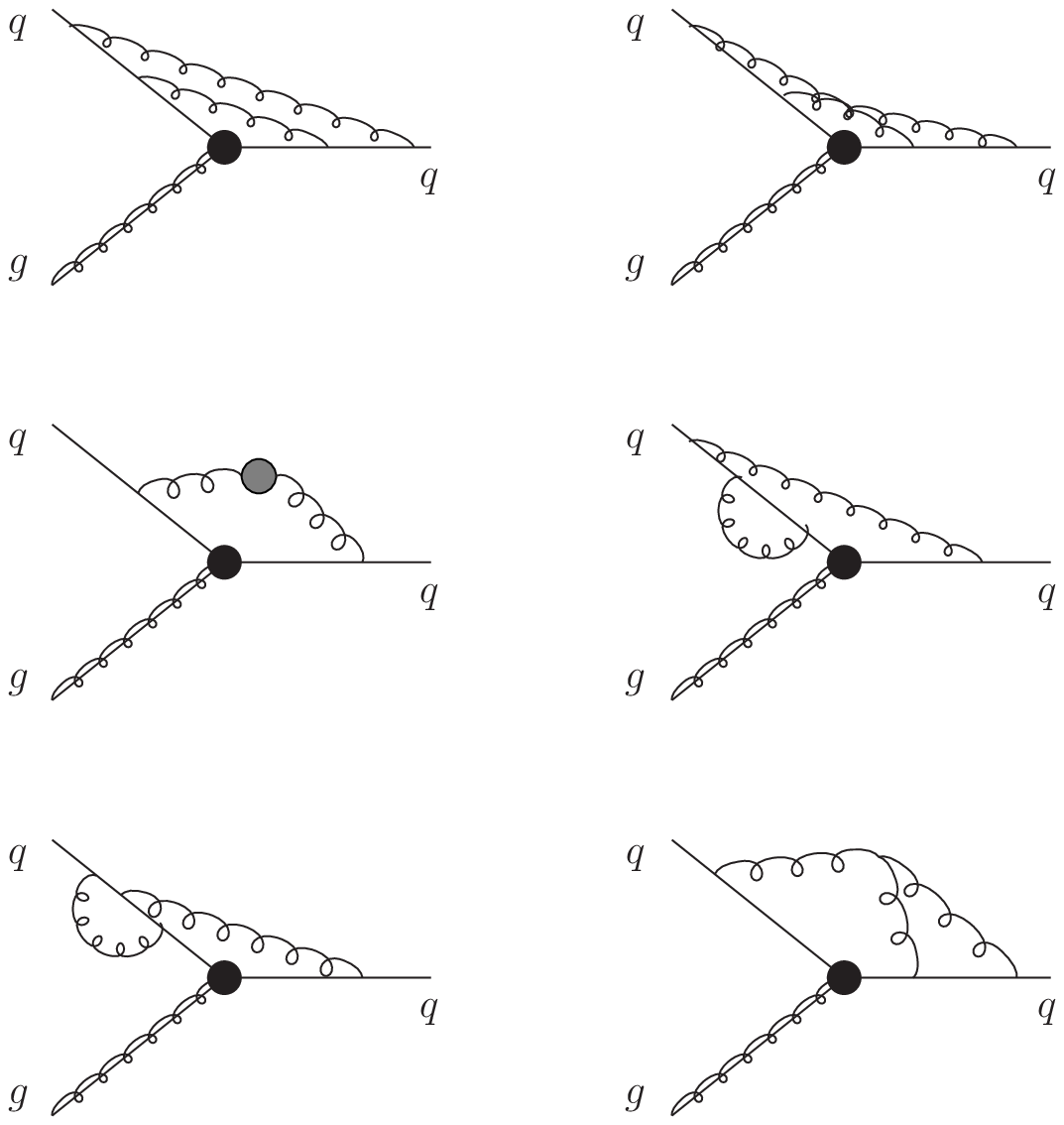}
\caption{Two-loop eikonal diagrams for $W$ or $Z$ production-Set 2.}
\label{W2loop2}
\end{figure}

\begin{figure}[h]
\centering
\includegraphics[width=80mm]{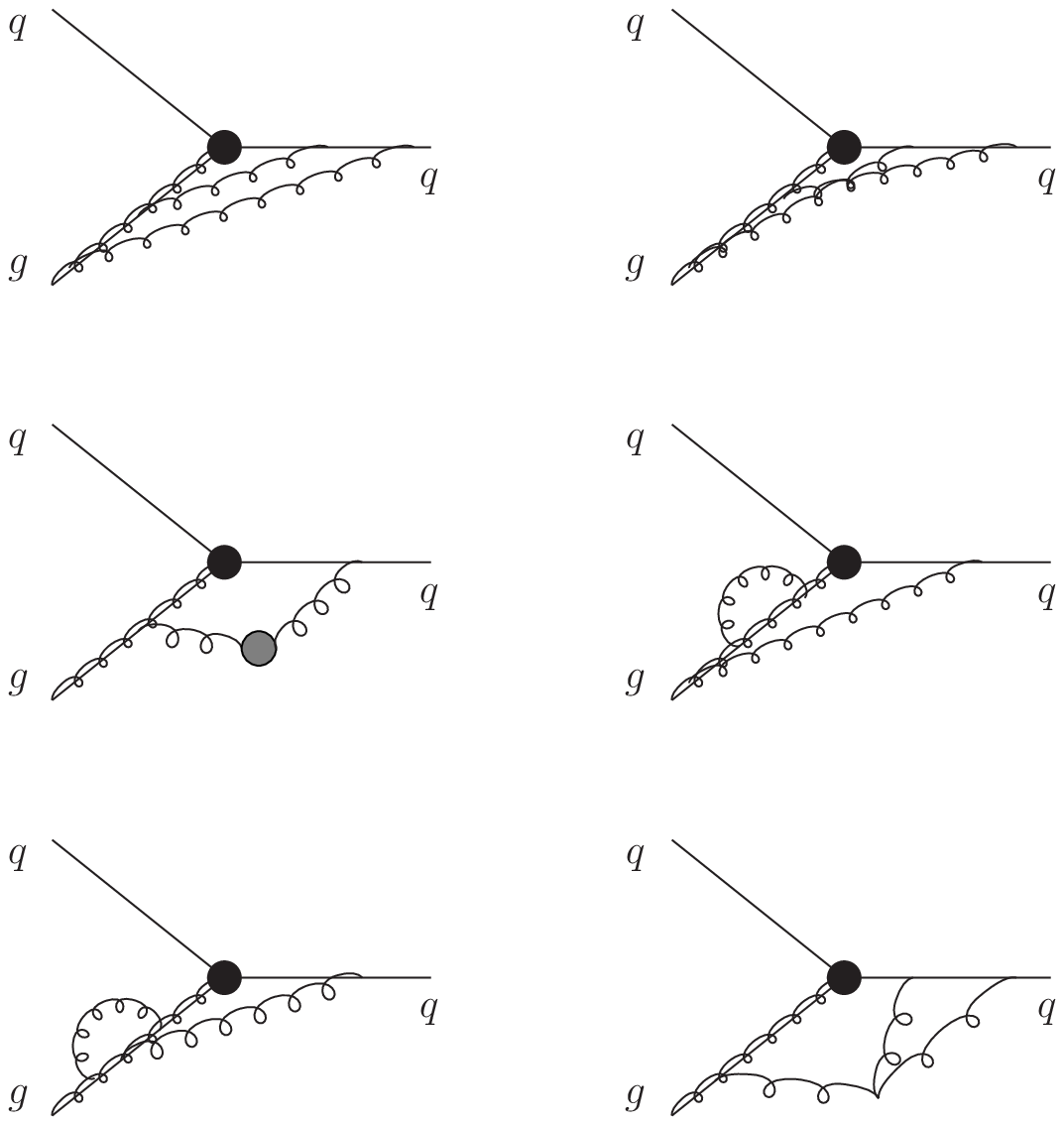}
\caption{Two-loop eikonal diagrams for $W$ or $Z$ production-Set 3.}
\label{W2loop3}
\end{figure}

We determine $\Gamma_S^{(2)}$ from the UV poles of two-loop 
dimensionally regularized integrals for the eikonal diagrams shown in 
Figs. \ref{W2loop1}, \ref{W2loop2}, \ref{W2loop3}.

For $qg\rightarrow Wq$ (or $qg\rightarrow Z q$)  
the one-loop soft anomalous dimension is 
$$
\Gamma_{S,\, qg\rightarrow Wq}^{(1)}=C_F \ln\left(\frac{-u}{s}\right)
+\frac{C_A}{2} \ln\left(\frac{t}{u}\right)
$$
and the two-loop soft anomalous dimension is 
$$
\Gamma_{S,\, qg \rightarrow Wq}^{(2)}=\frac{K}{2} \Gamma_{S,\, qg \rightarrow Wq}^{(1)} \, ,
$$
where $K=C_A(67/18-\zeta_2)-5n_f/9$.

For $q {\bar q}\rightarrow Wg$ (or $q {\bar q}\rightarrow Z g$) the 
corresponding results are 
$$
\Gamma_{S,\, q{\bar q}\rightarrow Wg}^{(1)}=\frac{C_A}{2} \ln\left(\frac{tu}{s^2}\right)$$
and 
$$
\Gamma_{S,\, q{\bar q} \rightarrow Wg}^{(2)}=\frac{K}{2} \Gamma_{S,\, q{\bar q} \rightarrow Wg}^{(1)} \, . $$

\section{$W$ production at the Tevatron and the LHC}

\begin{figure}[h]
\centering
\includegraphics[width=100mm]{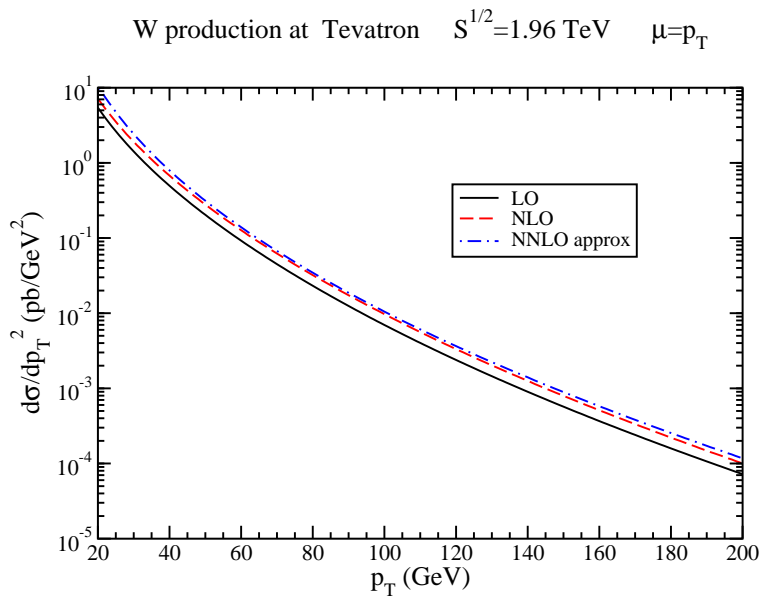}
\caption{NNLO approximate $p_T$ distribution at the Tevatron.}
\label{Wnnlotev}
\end{figure}

In Fig. \ref{Wnnlotev} we plot LO, NLO, and approximate NNLO 
results for the $W$-boson $p_T$ distribution at the Tevatron.
We have set $\mu=\mu_F=\mu_R=p_T$.
We note that the NLO corrections are large and the NNLO approximate 
corrections are significant.

\begin{figure}[h]
\centering
\includegraphics[width=100mm]{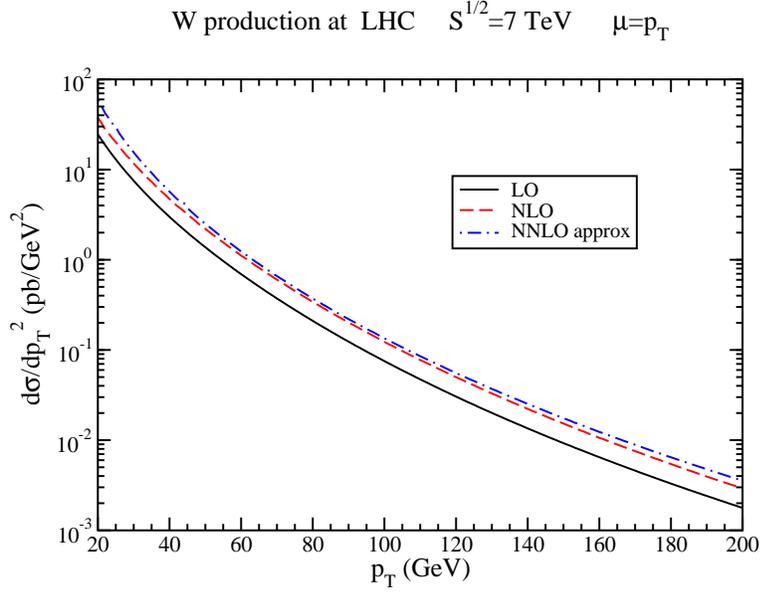}
\caption{NNLO approximate $p_T$ distribution at the LHC at 7 TeV.}
\label{Wnnlo7lhc}
\end{figure}

In Fig. \ref{Wnnlo7lhc} we plot LO, NLO, and approximate NNLO 
results for the $W$-boson $p_T$ distribution at the LHC at 7 TeV.
We note that the NLO corrections are even larger than at the Tevatron and 
the NNLO approximate corrections are again significant.

\begin{figure}[h]
\centering
\includegraphics[width=100mm]{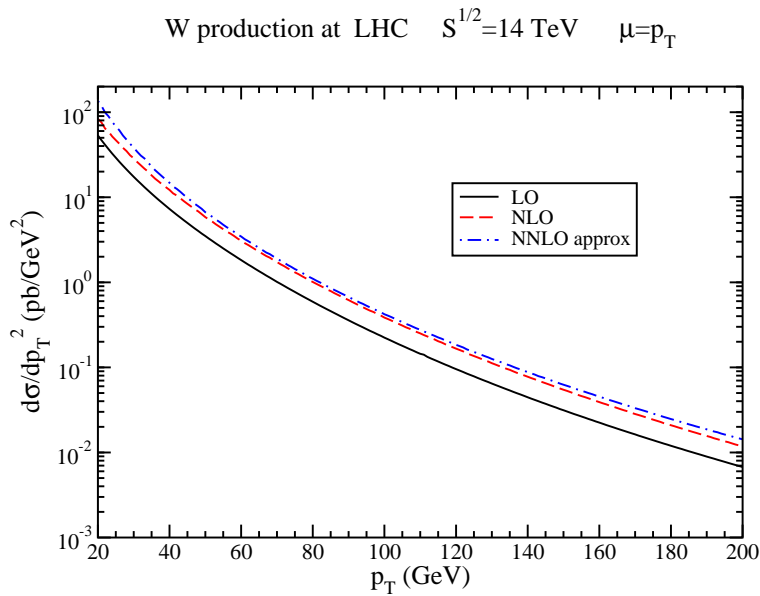}
\caption{NNLO approximate $p_T$ distribution at the LHC at 14 TeV.}
\label{Wnnlo14lhc}
\end{figure}

The corresponding results at 14 TeV energy at the LHC are shown in 
Fig. \ref{Wnnlo14lhc}, with similar conclusions for the size of the 
higher-order corrections.

\section{Summary}
 
We have presented new theoretical results for $W$ and $Z$ production 
at large $p_T$. 
We have discussed LO and NLO results and identified the 
soft-gluon threshold corrections as an important set of corrections.
We employed two-loop resummation to calculate
NNLO threshold corrections.
These corrections are significant and must be included for greater 
theoretical accuracy.
New results for the approximate NNLO $p_T$ distribution in $W$ production 
at the Tevatron and LHC were presented. More work is under way.

%%%%%%%%%%%%%%%%%%%%%%%%%%%%%%%%%%
\begin{acknowledgments}
This work was supported by the National Science Foundation under 
Grant No. PHY 0855421.
\end{acknowledgments}

\bigskip % extra skip inserted
%\begin{thebibliography}{9}   % Use for  1-9  references

\end{document}